# De la Extensión a la Investigación: Como La Robótica Estimula el Interés Académico en Estudiantes de Grado

Flores Gabriela, luisa.flores@estudiantes.utec.edu.uy[1]
Mazondo Ahilen, ahilen.mazondo@estudiantes.utec.edu.uy[1]
Moraes Pablo, pablo.moraes@utec.edu.uy[1]
Sodre Hiago, hiago.sodre@utec.edu.uy[1]
Peters Christopher, qristopherp@gmail.com[2]
Saravia Victoria, victoria.saravia@utec.edu.uy[1]
Da Silva Angel, angel.dasilva@estudiantes.utec.edu.uy[1]
Fernández Santiago, santiago.fernandez@estudiantes.utec.edu.uy[1]
de Vargas Bruna, bruna.devargas@utec.edu.uy[1]
Kelbouscas André, andre.dasilva@utec.edu.uy[1]
Grando Ricardo, ricardo.bedin@utec.edu.uy[1]
Assunção Nathalie, nathalie.assuncao@utec.edu.uy[1]

[1]Universidad Tecnológica del Uruguay
[2]Ostfalia University of Applied Sciences

***Abstract:*** *This research examines the impact of robotics groups in higher education, focusing on how these activities influence the development of transversal skills and academic motivation. While robotics goes beyond just technical knowledge, participation in these groups has been observed to significantly improve skills such as teamwork, creativity, and problem-solving. The study, conducted with the UruBots group, shows that students involved in robotics not only reinforce their theoretical knowledge but also increase their interest in research and academic commitment. These results highlight the potential of educational robotics to transform the learning experience by promoting active and collaborative learning. This work lays the groundwork for future research on how robotics can continue to enhance higher education and motivate students in their academic and professional careers.*

***Resumen:*** *Esta investigación examina el impacto de los grupos de robótica en la educación superior, enfocándose en cómo estas actividades influyen en el desarrollo de habilidades transversales y la motivación académica. Aunque la robótica no se limita solo a conocimientos técnicos, se observó que la participación en estos grupos mejora significativamente habilidades como el trabajo en equipo, la creatividad y la resolución de problemas. El estudio, realizado con el grupo UruBots, muestra que los estudiantes involucrados en robótica no solo refuerzan sus conocimientos teóricos, sino que también aumentan su interés en la investigación y su compromiso académico. Estos resultados destacan el potencial de la robótica educativa para transformar la experiencia de aprendizaje, promoviendo un aprendizaje activo y colaborativo. Este trabajo sienta las bases para futuras investigaciones sobre cómo la robótica puede seguir mejorando la educación superior y motivando a los estudiantes en sus carreras académicas y profesionales.*

***Palabras clave:*** *Estudiantes, Robótica, Interés Académico, Motivación Académica, Investigación.*

## 1. INTRODUCCIÓN

En los últimos años, la robótica ha emergido como una disciplina fundamental en la educación superior, no solo como una especialización técnica, sino como una herramienta clave para fomentar el interés académico y desarrollar habilidades transversales en los estudiantes de grado (Gratani & Giannandrea, 2022; García & López, 2020). En las universidades, los grupos de robótica juegan un papel crucial en la enseñanza práctica de la ingeniería, despiertan la curiosidad y el interés por la investigación, y brindan a los estudiantes la oportunidad de aplicar sus conocimientos en proyectos complejos y en la creación de nuevas soluciones tecnológicas. La robótica en la educación va más allá de la comprensión de principios mecánicos y electrónicos; también promueve habilidades esenciales como el pensamiento crítico, la resolución de problemas y la capacidad de trabajar en equipo. Estas competencias son cruciales en un entorno cada vez más interdisciplinario y complejo, donde la integración de conocimientos de diversas áreas es vital para el éxito académico y profesional. Además, la participación en grupos de robótica ofrece a los estudiantes una plataforma para experimentar con tecnologías emergentes, como la inteligencia artificial y el aprendizaje automático, ampliando así su visión sobre las aplicaciones de la robótica en distintos campos.

El objetivo de este artículo era identificar las habilidades transversales que los estudiantes de grado logran construir y desarrollar a través de su participación en grupos de robótica y explorar cómo estas actividades actúan como herramientas complementarias en su formación integral, además de apoyar la motivación para la permanencia en la carrera universitaria. Se examinó cómo la robótica puede transformar la experiencia educativa al promover un aprendizaje activo y colaborativo, y cómo este enfoque puede motivar a los estudiantes a involucrarse más profundamente en la investigación académica. La experiencia práctica proporcionada por estos grupos complementa la formación teórica, lo que puede aumentar la retención del conocimiento y su aplicación efectiva en situaciones reales. El artículo se organiza en las siguientes secciones: Introducción, que establece el contexto del estudio; Referencial Teórico, que revisa estudios previos sobre la robótica en la educación; Metodología, que describe el diseño del estudio realizado con estudiantes de grado; Resultados y Discusión, donde se presentan y analizan los hallazgos del estudio; y, finalmente, Conclusiones, que resumen los puntos clave y ofrecen recomendaciones para futuras investigaciones.

## 2. REFERENCIAL TEÓRICO

### 2.1 Robótica Educacional

La robótica educativa ha sido un campo de creciente interés tanto en la investigación académica como en la práctica pedagógica. Numerosos estudios han demostrado que la incorporación de la robótica en el entorno educativo no solo mejora el aprendizaje de conceptos técnicos, sino que también promueve el desarrollo de habilidades blandas, como la resolución de problemas, el trabajo en equipo y la creatividad (Bers et al., 2018; Eguchi, 2014). Un aspecto clave que emerge de la literatura es cómo los grupos de robótica en las universidades actúan como un catalizador para que los estudiantes de grado se interesen por la investigación. Según Benitti (2012), la participación activa en proyectos de robótica permite a los estudiantes aplicar conocimientos teóricos en situaciones prácticas, lo que no solo refuerza su comprensión académica, sino que también despierta un interés por la investigación y el desarrollo de nuevas tecnologías

Además, estudios como los de González et al. (2021) han señalado que los estudiantes que participan en grupos de robótica suelen demostrar un mayor compromiso con su carrera, lo que a su vez se traduce en un mejor rendimiento académico y una mayor propensión a involucrarse en proyectos de investigación durante y después de su formación universitaria. Por último, es importante destacar el papel de la robótica como un puente hacia la innovación. La robótica no solo facilita la aplicación de conceptos aprendidos en clase, sino que también impulsa a los estudiantes a explorar nuevas ideas, experimentar con tecnologías emergentes y desarrollar soluciones creativas para problemas complejos (Mubin et al., 2013). Estos factores hacen que la robótica sea un campo especialmente relevante para fomentar el interés académico y la investigación en estudiantes de grado.

Figura 1. Interacción entre un robot humanoide y niños.

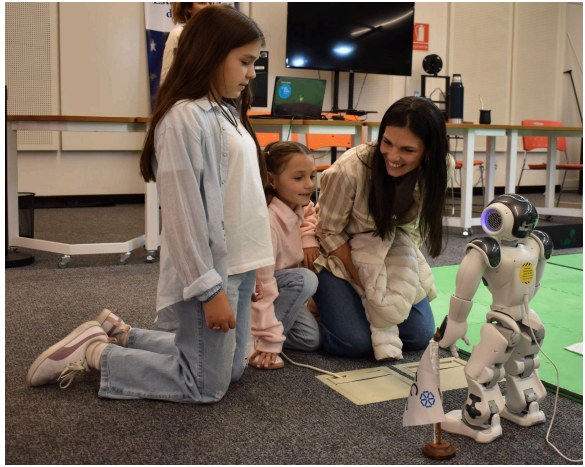

**2.2 Grupos de Robótica**

**Cañamero & Aylet (2020)** explora cómo la participación en el Desafío FIRST® LEGO® League impacta positivamente en las actitudes de los estudiantes hacia STEM, su aprendizaje y el desarrollo de habilidades del siglo XXI, como el pensamiento crítico y la resolución de problemas. Los hallazgos indican que estas competencias se fortalecen significativamente gracias a la participación activa en competencias de robótica.

**Karim (2013)** explora el impacto de la robótica educativa en el aprendizaje de STEM (ciencia, tecnología, ingeniería y matemáticas), las actitudes de los estudiantes y las habilidades necesarias en el lugar de trabajo. A través de una revisión exhaustiva, el autor examina cómo la integración de robots en el entorno educativo no solo mejora el conocimiento técnico, sino que también fomenta habilidades blandas esenciales como la colaboración, la resolución de problemas y la adaptabilidad. Este trabajo subraya la importancia de preparar a los estudiantes para los desafíos del siglo XXI mediante el uso de herramientas tecnológicas innovadoras como la robótica educativa.

**Jones & Wang (2022)** realiza una revisión sistemática sobre el uso de la robótica en la educación, destacando su efectividad para mejorar los resultados de aprendizaje en STEM. Se enfatiza cómo la robótica puede involucrar a los estudiantes de manera más profunda en su proceso educativo, promoviendo la colaboración y el pensamiento crítico en el aula.

**Anggoro & Wijaya (2020)** investiga la robótica como una herramienta eficaz en la educación STEM, mostrando que su implementación reduce la brecha entre teoría y práctica. Se destaca que la robótica no solo facilita la comprensión de conceptos STEM, sino que también fomenta habilidades importantes como la resolución de problemas y el trabajo en equipo.

Estos estudios subrayan el papel fundamental de la robótica educativa en el desarrollo de competencias transversales, lo cual se alinea con los objetivos del presente trabajo de investigación. La evidencia proporcionada destaca cómo la participación en grupos de robótica no solo fortalece el conocimiento técnico y las competencias en Science, Technology, Engineering and Mathematics, sino que también impulsa habilidades cruciales en la actualidad, como el pensamiento crítico, la resolución de problemas, la colaboración y la adaptabilidad. La robótica se presenta como una herramienta eficaz para conectar la teoría con la práctica, promoviendo un aprendizaje más profundo y significativo basado en proyectos. Con base a esto, el estudio, por lo tanto, se apoya en estos hallazgos para explorar cómo la participación activa en grupos de robótica puede facilitar el desarrollo integral de los estudiantes, preparándolos para enfrentar los desafíos de un entorno cada vez más tecnológico e interconectado.

**3. METODOLOGÍA**

La investigación se desarrolló mediante un estudio de caso enfocado en el grupo de robótica UruBots, en el cual participaron 23 estudiantes. Como instrumento de recolección de datos se diseñó y aplicó una encuesta semiestructurada. Esta encuesta, implementada en formato virtual, permitió a los estudiantes compartir sus experiencias de manera anónima y confidencial. Consta de 20 preguntas objetivas, divididas en dos secciones: la primera mitad dedicada a la identificación de características demográficas y académicas de los participantes, y la segunda sección enfocada en la evaluación de aspectos de la robótica asociados al desempeño social y académico de los estudiantes.

Este estudio empleó un abordaje cuantitativo mediante la aplicación de un formulario semiestructurado para evaluar la influencia de la participación en grupos de robótica en el rendimiento académico y el interés en la investigación. De acuerdo con Yin (2015), el enfoque de estudio de caso permite obtener una comprensión profunda del fenómeno en un contexto real, lo cual fue clave para analizar el impacto del grupo de robótica UruBots. La figura 2, presenta el encabezado del cuestionario enviado al público objetivo.

Figura 2. Propuesta del formulario de evaluación.

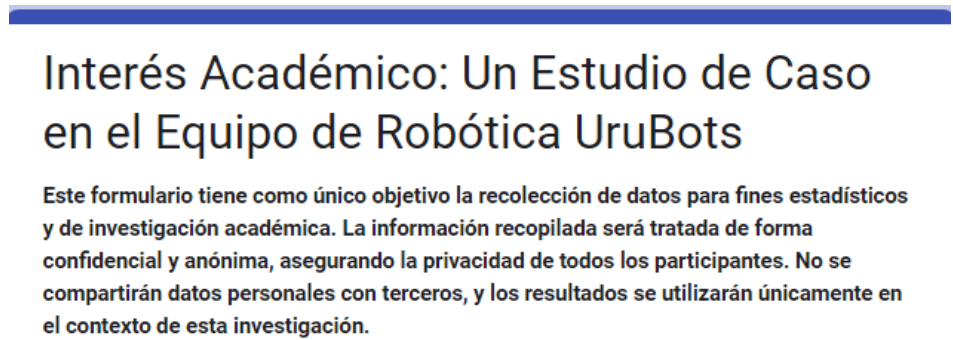

Con la evaluación de este enfoque aplicado a un equipo de robótica en cuestión, se buscó explorar una alta variedad de personas de distintas edades, nivel académico y distinto nivel de experiencia en el área de robótica. Se incluyeron personas que tienen algún vínculo con Urubots, lo que permitió obtener una perspectiva más completa sobre el impacto de la robótica en diferentes etapas de la carrera académica, destacando tanto los beneficios inmediatos como los de largo plazo.

## 4. RESULTADOS Y DISCUSIONES

Figura 3. Gráficos respuesta de Identificación del Grupo de Estudio

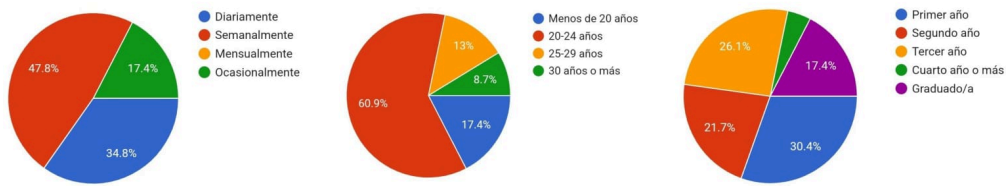

Con relación al perfil del público objetivo, el 60% (14 participantes) tiene entre 20 y 24 años, y el 87 % ha participado en competencias de robótica, dedicando de manera significativa su tiempo a estas actividades, ya sea diariamente o semanalmente.

El equipo es diverso, con miembros en diferentes etapas de su carrera, lo que aporta una variedad de niveles de experiencia. Como se observa en la figura 3, el 30,4% (7 participantes) están en su primer año, mientras que el 17,4% (4 participantes) ya son graduados. Además, se puede observar que hay estudiantes de segundo y tercer año.

Figura 4. Gráfico respuesta de "¿Qué habilidades blandas consideras que la robótica ha mejorado en ti?"

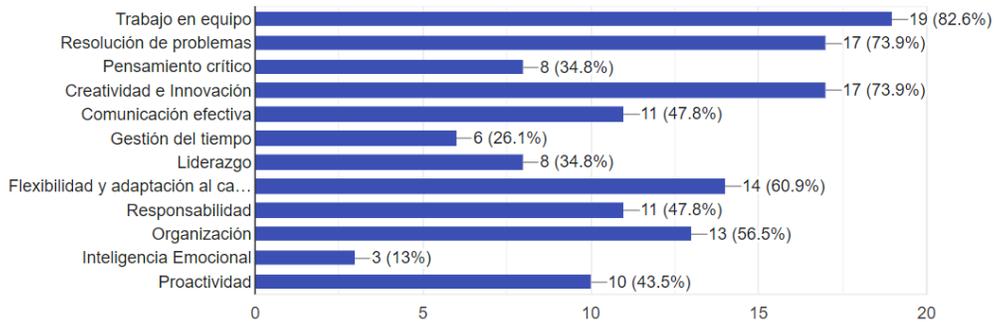

Se puede observar en la figura 4, que 20 de los participantes notó una mejora significativa en sus habilidades blandas, las cuales aplican no solo en su carrera académica sino también en su vida cotidiana. En la encuesta realizada, se incluyó la pregunta: "¿Qué habilidades blandas consideras que la robótica ha mejorado en ti?" Entre las respuestas más frecuentes destacan mejoras en el trabajo en equipo (82,6 %), creatividad e innovación (73,9 %) y resolución de problemas (73,9 %). Es importante relacionar que las habilidades dichas por los participantes están de acuerdo con las competencias valoradas por Young Global Leaders, como liderazgo, innovación y trabajo en equipo. Estas actividades en el ámbito de la robótica juegan un papel crucial en la motivación de los estudiantes, fomentando un mayor entusiasmo, siendo así un soporte en la permanencia de las carreras.

Figura 5. "¿La participación en proyectos de robótica ha contribuido a que sigas adelante con tu carrera?"

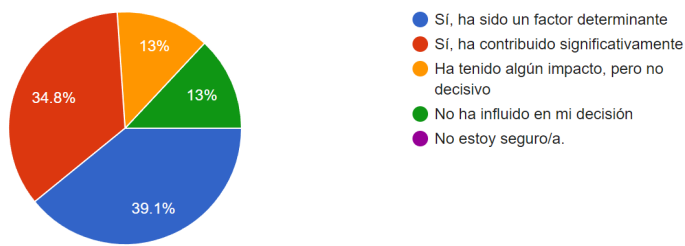

Como se ve en la figura 5, se muestra que el 86,9% (20 participantes) consideran que involucrarse en proyectos de robótica influye positivamente en sus carreras. Los participantes mostraron un incremento en su compromiso y una notable mejora en sus habilidades para trabajar en equipo, resolver problemas. Además, estos grupos de robótica ofrecieron una plataforma para experimentar con tecnologías emergentes, ampliando así el horizonte académico de los estudiantes y estimulándolos a involucrarse más en la investigación.

Con este trabajo fue posible extraer estadísticas importantes acerca de la influencia de trabajos en equipos de robótica a lo que relaciona al desarrollo personal de los estudiantes. Este enfoque permitirá identificar con mayor precisión los beneficios específicos en el desarrollo de habilidades blandas y su impacto en la formación integral de los estudiantes. A continuación se presentan las gráficas de resultados del cuestionario implementado.

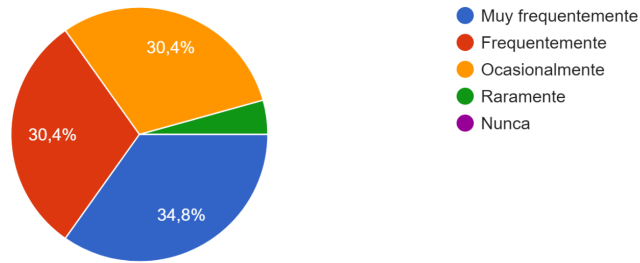

Figura 6. "¿Con qué frecuencia utilizas las habilidades aprendidas en robótica en tu vida diaria o en otros aspectos de tu educación?"

Se obtuvieron un total de 23 respuestas. Con base a lo que fue recolectado los resultados muestran puntos relevantes relacionados a la participación de los sujetos en grupos de robótica. Indicaron que dicha participación fomenta un mayor interés académico y mejora la retención de estudiantes en sus carreras.

## 5. CONCLUSIONES

Se espera ampliar la encuesta a estudiantes de diferentes universidades para obtener una visión más detallada sobre cómo la participación en grupos de robótica influye en el interés académico y la preparación para la investigación. La robótica educativa desempeña un papel crucial en la estimulación del interés académico, los grupos de robótica ofrecen una experiencia práctica que complementa la formación teórica, promoviendo un aprendizaje activo y colaborativo. Este estudio ha demostrado que la participación en grupos de robótica tiene un impacto significativo en aspectos como: en el interés académico y la preparación para la investigación de los estudiantes de grado.

Los resultados indican que los estudiantes involucrados en grupos de robótica no solo muestran un mayor compromiso con sus estudios, sino que también desarrollan habilidades blandas cruciales como el trabajo en equipo, la creatividad y la resolución de problemas. Estas habilidades, a su vez, pueden mejorar su rendimiento académico y su capacidad para enfrentar desafíos complejos.

La experiencia práctica proporcionada por los grupos de robótica complementa eficazmente la formación teórica, promoviendo un aprendizaje activo y colaborativo que resulta en una mayor motivación y entusiasmo hacia las carreras académicas y profesionales. En conclusión, los grupos de robótica representan una valiosa herramienta educativa que no solo mejora el aprendizaje práctico, sino que también impulsa la motivación y el compromiso de los estudiantes con sus carreras. Este estudio destaca la necesidad de implementar programas de formación en robótica en un mayor número de instituciones educativas y continuar investigando su impacto para maximizar sus beneficios educativos y profesionales. Es importante reconocer que la robótica no solo fortalece las habilidades técnicas, sino que también contribuye al desarrollo de habilidades blandas, como la comunicación, el trabajo en equipo y la creatividad. Al incluir la robótica en los procesos de enseñanza y aprendizaje, transforma la educación al proporcionar experiencias prácticas que promueven el pensamiento crítico y la capacidad de resolver problemas de manera innovadora.

## 6. REFERENCIAS